\documentstyle[12pt]{article} 
\newcommand{\be}{\begin{equation}}
\newcommand{\ee}{\end{equation}}

\newcommand{\ii}{\'{\i}}
\newcommand{\bt} { \begin{tabular} }
\newcommand{\et}{ \end{tabular} }
\newcommand{\bc} { \begin{center} }
\newcommand{\ec}{ \end{center} }
\newcommand{\mc}{ \multicolumn }
\newcommand{\f}{ \frac }

\newcommand{\la}{\label }
\newcommand{\bfi}{\begin{figure} }
\newcommand{\efi}{\end{figure} }

\newcommand{\btb} { \begin{table} }
\newcommand{\etb}{ \end{table} }
\begin{document}
\title{THE CRITICAL BEHAVIOUR OF THE SPIN-3/2 BLUME-CAPEL MODEL IN TWO DIMENSIONS}
\author{J. C. \ Xavier \ , \  F. C. \ Alcaraz \\
	 Departamento de F\'{\i}sica \\
	Universidade Federal de S\~ao Carlos \\
	 13565-905, S\~ao Carlos, SP, Brazil \\
\vspace{.05cm} \\
D. \ Pen\~a Lara $^*$\  and \  J. A. \ Plascak \\
	 Departamento de F\'{\i}sica \\
	Universidade Federal de Minas Gerais \\
	 30161-970, Belo Horizonte, MG, Brazil }
\date{PACS numbers : 64.60Kw, 64.60Cn, 64.60Fr }
\maketitle
\vspace{0.2cm}
\begin{abstract}
The phase diagram of the spin-3/2 Blume-Capel model in two dimensions is 
explored by conventional finite-size scaling, conformal invariance and Monte
Carlo simulations. The model in its $\tau$-continuum Hamiltonian version is
also considered and compared with others spin-3/2 quantum chains. Our
results  indicate that differently from the standard spin-1 Blume-Capel
model there is no multicritical point along the order-disorder
transition line. This is in qualitative agreement with mean field prediction
but in disagreement with previous approximate renormalization group
calculations.
\end{abstract}

$^*$ Present Address: Departamento de Fisica, Universidad del Valle,
AA 25360 - Cali - Colombia.

\section{Introduction}
The Spin-S Blume-Capel Model is a generalization of the standard Ising model
with dynamics described by the Hamiltonian
\be 
H=-J\sum_{<i,j>}s_is_j+D\sum_{i}s_i^2,
\la{hbc}
\ee
where the first sum runs over all nearest neighbor and the spin variables $%
s_i$ assume values $-S,-S+1,...,S$, associated with each site $i$ of the
lattice. In Eq.(\ref{hbc}) $J$ is the exchange coupling and $D$ is a single
spin anisotropy parameter. In the case where $S=1$ this Hamiltonian was
proposed originally by Blume and Capel \cite{s1mf} for treating magnetic
systems and has also been used in describing  
$^3$He-$^4$He mixtures \cite{beg}. This $S=1$ model was
 studied by a variety of methods
such as mean field \cite{s1mf}, two-spin cluster \cite{s1sc}, variational
methods \cite{s1vm}, constant coupling approximation \cite{s1cc}, Monte
Carlo simulations \cite{s1mc,s1tri}, finite-size-scaling on its transfer matrix 
\cite{s1fss} or quantum Hamiltonian \cite{s1c.7,drugo} and renormalization 
group methods \cite{s1rg,srg}. It is well established 
that for dimension $d\ge 2$ the $S=1$
model presents a phase diagram with ordered ferromagnetic and disordered
paramagnetic phases separated by a transition line which changes from a
second-order character (Ising type) to a first-order one at a tricritical
point. More specifically in two dimensions the machinery coming from
conformal invariance \cite{ci1,ci2} indicates that at this tricritical point
the long-range fluctuations are governed by a conformal field theory with
central charge $c=7/10$ \cite{s1c.7,drugo}. In this case all the critical
exponents and the whole operator content of the model are obtained \cite
{drugo}.

For values of spin $S>1$ however the situation is quite unclear with fewer
results and contradictions among them. The mean field calculation \cite
{s1.5mf} (see also \cite{salinas}) predicts different phase diagrams 
for integer or half-odd-integer
spins. For spin $S=3/2$ it gives a second order phase transition with no
tricritical point and a separated first-order transition line which ends up
in an isolated multicritical point. In contradiction with these results a
calculation \cite{srg} based on a renormalization group introduced in 
\cite{rgin} give us for the two dimensional model a unique first-order
transition line at low temperature which ends up in the second-order
transition line at a tetracritical point. Similar results are also obtained 
 by using a Migdal-Kadanoff real space renormalization group
approach \cite{s1.5mk}.

Motivated by these contradictions we decided to study extensively the
two-dimensional $S=3/2$ model by using conformal invariance and finite-size
scaling, since these methods give us in the $S=1$ model the most precise and
conclusive results. In order to supplement our conclusions, in the 
$S=3/2$ model, we also perform
Monte Carlo simulations in the region of the phase diagram where
multicritical behavior might occur.

The layout of this paper is as follows. In section 2 the transfer matrix of
the model and the relations used in our finite-size studies are presented. In
section 3 we discuss initially our results for the $S=1$ model. Although
the phase diagram is well established we believe that our numerical
estimate for the tricritical point is the most precise in the present
literature. In section 4 our results for the spin 3/2 is presented and
discussed. In section 5 we study the spin-3/2 model in its $\tau $%
-continuum formulation and compare its quantum Hamiltonian with another
spin-3/2 quantum chain which is known to present multicritical point along
the order-disorder phase transition line. In section
6 we close our paper with a summary and conclusions of our main results, and
finally in appendix A the details and methods of our Monte Carlo simulations
for the spin-3/2 model are explained. 
\section{Transfer Matrix and Conformal Invariance Relations} 
The row-to-row transfer matrix $\hat T$ of the Hamiltonian (\ref{hbc}) in a
square lattice, with horizontal width $N$ has dimension $(2S+1)^N\times
(2S+1)^N$. Its coefficients $<s_1^{\prime },...,s_N^{\prime }|\hat
T|s_1,...,s_N>$ are the Boltzmann weights generated by the spin
configurations $\{s_1,...,s_N\}$ and $\{s_1^{\prime },...,s_N^{\prime }\}$
of adjacent rows. In general, if we
consider (\ref{hbc}) with periodic or free boundary conditions in the
horizontal direction, the matrix $\hat T$ will have all  its
elements non-zero, being a dense matrix. This difficulty restricts the numerical
study for very small lattices. A possible way to reach larger lattices, with
reasonable computational effort, is to consider the model in its $\tau $%
-continuum limit \cite{cl}. In this case after a high anisotropic limit of
its couplings in the vertical and horizontal directions we replace the
transfer matrix by a simpler effective quantum Hamiltonian which is a sparse
matrix. Most of the finite-size scaling studies exploring conformal
invariance, for general models were done in this formulation. However in
such approach the calculation of the transition lines, in terms of the original
coupling (isotropic) is not possible. Another possibility, which is suitable
for our purposes, is to impose helical boundary conditions along the
horizontal direction. In this case the transfer matrix can be written as
\be
\hat T=\hat t^N,
\la{tm}
\ee
where $\hat t$ is the transfer submatrix with elements given by
\be
<s_1,...,s_N|\hat t|s_1',...,s_N'>=\exp
\left(t^{-1} s_1(s_2+s_N')-dt^{-1} s_1^2 \right)
\prod_{j=1}^{N-1}\delta_{s_{j+1},s_j'},
\la{sm}
\ee
with $t=k_BT/J$ ($k_B$ is the Boltzmann constant) and $d=D/J$. We also see
from (\ref{tm}) and (\ref{sm}) that although $\hat T$ is a dense matrix $%
\hat t$ is a sparse one, having only a fraction $f=\exp \left(
-(N-1)\ln (2S+1)\right) $ of non-zero elements. The Hamiltonian (\ref{hbc}),
like the standard Ising model has a global $Z(2)$ symmetry. The helical
boundary condition does not break this symmetry, which translates in the
commutation of $\hat t$ with the parity operator
$$
\hat P=\prod_{i=1}^{N}\hat R_{l},
$$
where $\hat R_l=1\otimes 1\otimes \cdots 1\otimes \hat R\otimes 1\cdots
\otimes 1$, and $\hat R$, located at $lth$ position in the product, are $%
(2S+1)\times (2S+1)$ matrices given by
$$
\hat R=\sum_{s=-S}^{S}|s><-s|.
$$
Consequently in the basis where $\hat P$ is diagonal we can separate the
vector space associated to $\hat T$ into two disjoint sectors labelled by the 
eigenvalues $p=\pm 1$ of the parity operator.

Let us for convenience introduce the Hamiltonian matrix $\hat H=-\ln (\hat
T)=-N\ln (\hat t)$, and denote by the $e_{l,p}$ the $lth$ lowest ($l=1,2,...)
$ eigenvalue in the sector with parity $p$. The eigenenergies of the ground
state and first excited state are given by $e_{1,1}$ and $e_{1,-1}$,
respectively. These two leading energies, and consequently the mass gap 
$G_N(t,d)=e_{1,-1}-e_{1,1}$, for a system of width $N$ are real and
 can be obtained for
relative large $N$ by a direct application of the power method \cite{livro}.
 The standard finite-size scaling \cite{fss} gives an estimate of the phase transition
curve $t_c=t_c(d)$. This curve is evaluated by extrapolations to the bulk
limit ($N\rightarrow \infty $) of sequences $t_c(d,N)$ obtained by solving
\be
G_N(t_c)N=G_{N+1}(t_c )(N+1)\hspace*{1cm}N=2,3,...
\hspace{.15cm}.
\la{fss}
\ee
An heuristic method, which was proved to be effective in obtaining
multicritical points in earlier works \cite{s1c.7,tripoint,malvezzi} is to
simultaneously solve (\ref{fss}) for three different lattice sizes
\be
G_N(t_c)N=G_{N+1}(t_c )(N+1)=G_{N+2}(t_c)(N+2)\hspace*{1cm}N=2,3,...
\hspace{.15cm}.
\la{tripoint}
\ee
Once the transition curve is estimated, in the region of continuous phase
transition we expect the model to be conformal invariant. This symmetry
allows us to infer the critical properties from the finite-size corrections
of the eigenspectrum at $t_c$ \cite{ci1,ci2}. The appropriate relations which
give the scaling dimensions and conformal anomaly in the case of periodic or
free boundary conditions are well known in the literature \cite{ci1,ci2}. In the
present case, where we impose helical boundary conditions, these relations are
slightly different from the periodic case since translational invariance 
is lost in the horizontal direction. The conformal anomaly $c$, which 
labels the universality class can be calculated from the large-$N$ behavior 
of the ground-state energy
\be
\frac{e_{1,1}(N)}{N}= \epsilon_{\infty} - \frac{\pi c v_{s}}{6N^{2}} 
+o(N^{-2}),
\la{ano}
\ee
where $\epsilon _\infty $ is the ground-state energy, per site, in the bulk
limit and $v_s$, which is unity in our transfer matrix calculations, is the
sound velocity. The scaling dimensions of operators governing the critical
fluctuations (related to critical exponents) are evaluated from the finite-$N
$ corrections of the excited states. For each primary operator, with
dimension $x_\phi $, in the operator algebra of the system, there exists an
infinite tower of eigenstates of $\hat H$ whose energy $e_{m,m^{\prime
}}^\phi $ are given by
\be
\Re\left( e_{m,m'}^{\phi}(N)\right)=e_{1,1}+\frac{ 2\pi v_s }{ N }(x_{\phi}+m+m') + o(N^{-1}),
\la{dim}
\ee
where $m,m^{\prime }=0,1,\dots $ . The real part of the energies appears in (%
\ref{dim}) since in general they are complex due to the non-hermicity of $%
\hat H$. The multiplicity of energies with the same real part $\Re \left(
e_{m,m^{\prime }}^\phi (N)\right) $ are given in terms of the product of two
Virasoro algebras, as in the  periodic case. We tested extensively the
relations (\ref{ano}) and (\ref{dim}) for several excited eigenenergies in
the case of the Ising model with helical boundary conditions.
\section{The spin-1 Model}
In order to test our numerical methods and the conformal invariant relations
(\ref{ano}) and (\ref{dim}), we consider initially the Blume-Capel model
with spin $S=1$. Earlier finite-size studies that explores the conformal
invariance of this model were done in its $\tau $-continuum formulation \cite
{s1c.7,drugo}. In figure 1 we show the extrapolated ($N\rightarrow \infty $)
transition curve $t_c=t_c(d)$ (continuum line in the figure) obtained by
solving Eq.\ (\ref{fss}) for lattice sizes up to $N=12$. Note that for $d\ll -1
$ only configurations where $s_i=\pm 1$ is allowed and we have an effective
Ising model with $t_c=2/\ln (\sqrt{2}+1)=2.269...$.

The expected tricritical point of the model is obtained by extrapolating the
sequences obtained by solving Eq.\ (\ref{tripoint}). Our estimate for this
point is $t_t=0.609(4)$ and $d_t=1.965(5)$ (we solved (\ref{tripoint}) for
lattice sizes up to $N=10$), in agreement with previous Monte Carlo 
\cite{s1tri} and finite-size calculations \cite{s1fss}. In \cite{s1fss}
these estimates were done from data of lattice sizes up to $N=8$. The
extrapolations along this paper were obtained by using the $\epsilon $%
-alternated VBS approximation \cite{vbs}. The errors are estimated from the
stability of the extrapolations and are in the last digit. 
In figure 1 we show in a large scale the
region where the tricritical point is located. In this figure this point is
the end point of the continuum curve.

Strictly speaking the relation (\ref{fss}) only gives continuous phase
transitions. At the first-order line, in the spin-1 model, we have the
coexistence of three phases: two ordered ferromagnetic ($<s_i>\ne 0$) and
one disordered phase ($<s_i>=0$). Consequently the gaps related with the
three lowest eigenvalues vanishes exponentially as the lattice size
increases. A possible finite-size estimate for the first-order transition line,
 which we test in this paper, is
obtained by the following procedure. For a given lattice size $N$ we calculate
the points where the gap corresponding to the third eigenvalue has its minimum
value. The extrapolation $N\rightarrow \infty $ of these points give us our
estimate for the first-order transition line. The dotted line shown in Fig.
1 was obtained by this procedure. As we can see in this figure, the estimated 
first-order transition line (dotted line)  finishes at the tricritical
point.
 
In the critical regions of the phase transition line (continuum curve) the
conformal anomaly and the scaling dimensions can be calculated exploring the
conformal invariant relations (\ref{ano}) and (\ref{dim}). From Eq.\ (\ref{ano})
a possible way to extract $c$ is by extrapolating the sequence
\be
c^{N,N+1}=\f{6}{\pi}
\left( \f{e_{1,1}(N+1)}{N+1}-\f{e_{1,1}(N)}{N}
 \right)
\left( \f{1}{N^2}-\f{1}{(N+1)^2} \right)^{-1},
\la{anoest}
\ee
calculated at $t_c(d)$. Examples of such sequences together with the
extrapolated results for the spin-1 model, are shown in the first four
columns of table 1. We clearly see from this table the expected results, i.
e., an Ising behavior with conformal anomaly $c=1/2$ in the left of tricritical
point (continuum curve) and $c=7/10$ at the tricritical point.

From Eq. (\ref{dim}) the scaling dimensions $x(n,p)$ related to the
$nth$ ($n=1,2...$) energy in the sector with parity $p$ can be obtained by extrapolating the sequence
\be
x^N(n,p)=\f{ \left(\Re\left(e_{n,p}(N)\right)-e_{1,1}(N)
\right)N }{2\pi}.
\la{dimest}
\ee
The smallest dimension $x(1,-1)$ corresponds to a $Z(2)$-order parameter. 
Some of the related finite-size sequences are presented in the first four
 columns of the table 2 for the spin-1 model. As expect this dimension 
has the value 1/8 for the critical points in the Ising universality class
 and  3/40=0.075 at the tricritical point.
The conformal towers at tricritical point was obtained previously by using 
the $\tau $-continuum Hamiltonian formulation of the model \cite{s1c.7,drugo}.

In order to compare our estimate for the tricritical point with the 
previous results we have also calculated the conformal anomaly  and 
scaling dimensions by using the tricritical point estimated 
 in Refs. \cite{s1tri,s1fss}. 
These results show us that the extrapolated values of the conformal anomaly 
is similar to ours, however our estimate for the  scaling dimensions 
are much better than the values obtained at  their estimated tricritical point.
 For these reasons we believe that the tricritical point we found is a better 
estimate. 
\section{The spin-3/2 Model}
We now apply the numerical methods of section
2 in the controversial case of spin 3/2. In figure 2 we show the
extrapolated transition curve $t_c=t_c(d)$ (continuum line), obtained by
solving Eq.\ (\ref{fss}) for lattice sizes up to $N=10$. The limiting values $%
t_c=9/(2\ln (\sqrt{2}+1)=5.106...$ ($d\ll-1$) and $t_c=1/(2\ln (\sqrt{2}%
+1)=0.567...$($d\gg 1$) can be easily understood since in these limits the
model reduces to an Ising model. 

As we mentioned in the introduction there is a controversy in the literature
concerning the existence or not of a multicritical point along the
transition line. In order to clarify this point we try to obtain finite-size
estimates for this point by using relation (\ref{tripoint}). Our results
however show no consistent solutions of relations (\ref{tripoint}) for
lattice sizes up to $N=10$, which indicate the absence of a multicritical
point along the transition line. Another method, also used to locate
multicritical points \cite{tripoint}, is obtained from the simultaneous
crossing of two different gaps on a given pair of lattices (instead of three
lattices as in (\ref{fss})). Trying several different gaps we also did not
find, within this method, a multicritical point along the transition curve.

The first-order transition line was also estimated in a similar
 way as we did for
the spin-1 model in last section. Along this line we have the coexistence of
four ordered phases, with most of spins in state $s=\pm 3/2,1/2$ or $-1/2$
respectively. The finite-size estimator, for a given $N$, is given by the
points where the fourth gap has its minimum value, since in the bulk limit 
the four lowest eigenenergies will degenerate. The extrapolated curve
 ($N\rightarrow \infty $) is the dotted curve shown in Fig. 2. 
As depicted in a large scale in this figure, this first-order transition line 
does not touch the continuous transition line (differently from  the case $S=1$),
 in agreement with the topology  predicted for the phase diagram by mean field
 calculations \cite{s1.5mf}.

In order to supplement our results by an independent method we also did
Monte Carlo simulations for the spin-3/2 model. Details of the simulations
are presented in appendix A. The points for the phase diagram are the squares
shown in Fig. 2. Taking into account the scale of this figure the agreement
of the Monte Carlo simulations (squares) and the transfer matrix results
(circle) is good. These results indicate the absence of a multicritical
point along the continuous transition line. According to this scenario the
whole continuous phase-transition line should belong to the Ising universality
 class. In order to illustrate this point we present in the last four columns of
table 1, for some values of $d$, the finite-size sequences (\ref{anoest}) of
the conformal anomaly of the model. In the last four columns of table 2 we
also show some of finite-size sequences (\ref{dimest}) for the dimension 
$x(1,-1)$ of the order parameter. Again we obtain the expected results $1/8$
of the Ising-order parameter.

As usual in the region where the first-order phase transition takes place 
the finite-size sequences have slower convergence due to crossover effects. 
 This fact prevent us to get a good estimate for the
endpoint of the first order transition line. Our Monte Carlo simulations (
see appendix A) give an estimate $t_I=0.80$ and $d_I=1.96$, for this
point.
\section{The $\tau$-continuum quantum Hamiltonian of the spin-3/2 model and
other related quantum  chains}
Our results in the last section indicate that the spin-3/2 Blume-Capel
model does not have a multicritical point along the transition line that
separates the disordered phase from the ordered ones. Instead of working in
 the Euclidean version, we can also consider the model in its $\tau $-continuum
Hamiltonian formulation. This quantum Hamiltonian which is obtained by imposing
extreme anisotropic relations between its couplings in the horizontal and
vertical directions, can be derived straightforwardly (see \cite{cl}) for
example). For periodic boundary conditions it is given by $\hat H(b=1)$
where
\be
\hat H(b)=-\sum_{i}^{N}\hat S_i^z \hat S_{i+1}^z-
x_1(\hat S_i^z)^2-x_2\hat R_i(b),
\la{quant}
\ee
and $x_1$ and $x_2$ are coupling constants. The operators 
$\hat S_i$ and $\hat R_i(b)$ in the basis where $\hat S_i$ is
diagonal are given by
$$
\hat S_{i}^z=1\otimes 1\otimes \cdots 1\otimes S^z\otimes 1\cdots \otimes 1,
$$
$$
\hat R_{i}(b)=1\otimes 1\otimes \cdots 1\otimes R(b)\otimes 1\cdots \otimes 1,
$$
where the matrices $S^z$ and $R(b)$ are in the $ith$ position in the
product and are given by
\be
\begin{array}{cc}
R=
  \pmatrix{
  0   &1  &0   &0\cr
  1   &0  &b   &0\cr
  0   &b  &0   &1\cr
  0   &0  &1   &0\cr}
&
\hspace{.5cm}S^z=
  \pmatrix{
  3/2   &0  &0   &0\cr
  0   &1/2  &0   &0\cr
  0   &0  &-1/2   &0\cr
  0   &0  &0   &-3/2\cr}. \\
\end{array}
\la{rs}
\ee
%
%
%
Diagonalizing the Hamiltonian (\ref{quant}) for several lattices we try to
locate a multicritical point, by solving Eq.\ (\ref{fss}). In agreement with
the absence of multicritical behavior, we found no triple crossing for
lattice sizes $N>4$.

It is important to mention that, on the contrary to these results, the spin-3/2
Hamiltonian $\hat H(2/\sqrt{3})$, given by (\ref{quant}) is known to have
a tetracritical point along the transition line separating the disordered
and ordered line \cite{malvezzi,sen}. The matrices $S^z$ and $R(2/\sqrt{3})$%
, given in (\ref{rs}), are in this case the standard $S_z$-diagonal
representation of the spin-3/2 SU(2) matrices. This tetracritical point was
located \cite{malvezzi} by the same techniques we used in this paper and its
long-distance physics is related by a conformal field theory with central
charge $c=4/5$. Its is interesting to observe that such an small difference
in the Hamiltonians produce such differences in the critical behavior of the
Hamiltonians. In order to better see this fact numerically, we study the
Hamiltonian (\ref{quant}) with $b$ ranging from 1 to $2/\sqrt{3}$. We verify
that solutions of Eq.\ (7), for multicritical points  happens only for values 
of $b$ bigger than or very close to  $2/\sqrt{3}$. 

These results in the quantum Hamiltonian formulation indicate that
multicritical points along the order-disorder line can be generated only if
we introduce additional interactions in the spin-3/2 Blume-Capel model. The net
effect of the coupling $b$ in the quantum chain (\ref{quant}) is to control
the preference for neighboring spins $(s,s^{\prime })$ along the time
(vertical) direction to be in pairs (1/2,-1/2) or (-1/2,1/2). One of
possible ways to introduce this effect in the classical model is to add in
the spin-3/2 Blume-Capel (\ref{hbc}) Hamiltonian an extra term $V_{s,s^{\prime
}}(\alpha )=-\alpha (\delta _{s,1/2}\delta _{s^{\prime },-1/2}+\delta
_{s^{\prime },1/2}\delta _{s,-1/2})$, along all the links in the vertical.
The product $\beta \alpha $ will play a similar role as $b$ in (\ref{quant}%
). Studying numerically the transfer matrix of such classical Hamiltonian
 we found for $%
\alpha =1$ a multicritical point at $t_t=0.87(8)$ and $d_t=1.60(5)$.
Our calculation at this point are in favor of a value $c=4/5$ and $%
x(1,-1)=0.05$ for the conformal anomaly and the dimension of most relevant
order parameter, respectively. These are the predicted values for a minimal
conformal theory with $m=5$ \cite{sen}.
\section{Summary and Conclusion}
Our aim in this paper was the study of the critical properties of the
spin-3/2 Blume-Capel model by using the methods of finite-size scaling in
strip geometries, conformal invariance and Monte Carlo simulations. The
methods and useful relations for these calculation were presented in section
2 and in appendix A, and were tested for spin-1 Blume-Capel Model with a
precise evaluation of its phase diagram. Our results for the spin-3/2 help
to choose between the two existing contradictory predictions for its phase
diagram in ($t,d$)-parameter space. Mean-field calculations \cite
{s1.5mf} predict (scenario {\it {a}}) a phase diagram where there is no
multicritical point along the second-order phase transition line.
 In this scenario a
multicritical point will exist as an isolated end point of a first-order
 transition line. On the other hand approximate calculations based on
 renormalization- group methods in real space \cite{s1rg,srg} give us 
(scenario {\it {b}}) a multicritical point along the disorder-order transition
 line. This point being the end point  of the first-order
 transition line. Since the mean-field approximation is expected to be 
correct only for high dimensions and the other calculations predicting 
scenario $b$ were done directly in two dimensions we would expect, at least
 for two dimensions scenario $b$ as the correct one.
In parallel to these results there exist in the literature \cite{malvezzi,sen}
precise calculations on a related spin-3/2 quantum chain, which also
indicate a multicritical point along the transition line. Although these
calculations were done in a quantum Hamiltonian these results induce us in
favor of scenario {\it {b}}, since this quantum version chain $(\hat H(2/
\sqrt{3})$ in Eq. (\ref{quant})) has naively the same type of couplings as the
spin-3/2 Blume-Capel Model.

However, to our surprise, our results presented in section 4 
indicate the scenario {\it {a}} as the
correct one for the spin-3/2 Blume-Capel Model. In section 5 we
also analyze this last model in its $\tau $-continuum Hamiltonian
formulation $(\hat H(1)$ in Eq. (\ref{quant})) for the sake of comparison
with the above mentioned spin-3/2 quantum chain $(\hat H(2/\sqrt{3})$ in Eq. 
(\ref{quant})), which shows multicritical behavior along the order-disorder
transition line. Although these  Hamiltonians are very similar our finite-size
scaling studies indicate that these differences are enough to produce quite
distinct critical behavior. The results of section 5 also indicate that a
phase diagram for a spin-3/2 model like that in scenario {\it {b}} can be
produced by adding extra terms in the Blume-Capel model. 
\begin{center}
{\bf Acknowledgments}
\end{center}
 This work was supported in part by CNPq, CAPES, FAPESP, FAPEMIG
 and FINEP-Brazil.
\newpage
\appendix
\begin{center}
{\bf \large Appendix}
\end{center}
\section{Monte Carlo Simulations}
Our Monte Carlo simulations for the spin-3/2 Blume-Capel model were done
 on square lattices with $N^2$ sites and periodic
boundary conditions in all directions. We have used the standard single spin
flip algorithm \cite{binder} where the transition probability $P_{ij}$ from
a system configuration with energy $E_i$ (where a chosen spin is in a state $%
i$) to another configuration with energy $E_j$ (where this spin is now in a
state $j$) is
$$
P_{ij}  =  \frac{1}{2}[1 - \tanh{\frac{(\beta \Delta E_{ij})}{2}}]~,
$$
where $\Delta E_{ij} = E_j - E_i$. Since the spin variable can assume more than
two states we have chosen, at random, the state $j$ from the $(2S - 1)$
possible  distinct states from state $i$.
 
The thermodynamic quantities computed in these simulations
 are the magnetization per site
$$
m =  \langle M\rangle~,\label{eqn 2}
$$
the magnetic susceptibility
$$
\chi_m  =  \beta  N^2 \langle (M - m )^2\rangle~,\label{eqn 3}
$$
where
$$
M  =  \frac{\sum_i s_i}{N^2}~,\label{eqn 4}
$$
and the quadrupole magnetic susceptibility
$$
\chi _q =  \beta  N^2 \langle (q - \langle q\rangle )^2\rangle~,\label{eqn 5}
$$
where
$$
q  =  \frac{\sum_i s_i^2}{N^2}~.\label{eqn 6}
$$

Typical runnings have been made on a periodic 
lattice with $N=64$ spins and included $%
6.0\times 10^4$ iterations or Monte Carlo Steps (MCS) per spin of the
algorithm and the averages of the measured quantities were taken after
discarding the first $1.0\times 10^4$ MCS per spin. We have found that by
changing the lattice size from $N=64$ to $N=128$ the relevant measured
quantities did not change appreciably. For this reason, in most of the
computations we have taken $N=64$ in order to save computing time. In
addition, we have also calculated some Monte Carlo averages taking data
from: i) every MCS per spin; ii) every 10 MCS per spin; and iii) runnings
including $1.2\times 10^5$ MCS per spin with averages taken after discarding
the first $2.0\times 10^4$ MCS per spin. In all these runnings, no
significant differences have also been observed in the thermodynamic
quantities of interest when compared to the typical one previously stated in
the beginning of this paragraph. These results assure us that possible
correlations among different configurations, if any, are not relevant for
the computation of the magnetization and susceptibilities. Indeed, our
results, within the present numerical precision, are good enough to be
compared to other values obtained from different methods. For instance, at $%
d=0$ we have $t=3.32$ and in the spin-1/2 Ising limit $d\rightarrow \infty $
we have $t=0.58$ which is comparable to the exact value $t=0.57$.

The simulations for the phase diagram close to the isolated multiphase
critical point are shown in Fig. \ref{Fig2}. Second-order phase transitions
have been determined by the strong peak in the magnetic susceptibility $\chi
_m $, while first-order transitions  have been located through the
discontinuous behavior of the magnetization accompanied by a strong peak in
the quadrupole susceptibility $\chi _q $. A typical example is shown in
Figs. \ref{Fig3}, \ref{Fig4} for $d=1.998$ as a function of the reduced
temperature $t$. Such behavior of $\chi _q $ is useful for detecting 
first-order transitions when the discontinuity in the  magnetization of the
two ordered phases is very small. The discrepancy between the Monte
Carlo simulations and finite-size calculations for $d$ very close to $2$
reflects the large metastability of the ordered phases. For instance,
a metastable phase with $m=3/2$ will persist at low temperatures for
$d$ slightly above $2$ due to the ordering state starting configuration
($m=3/2$). In fact, by taking a random-state starting configuration and
sweeping up and down the temperature for constant $d\ge 2$ the only
stable phase found is the one having $m=1/2$ for $T=0$. The point
at $d=2.0$ and $t=0.4$ has not been simulated and is in Fig. 2 just
for the sake of clarity. 

The isolated multiphase critical point in the ordered phase is rather
difficult to be located with reasonable precision through Monte Carlo
simulations (an estimate can be given by $d_I=1.96$ and $t_I=0.80$).
Despite of that, we have carefully swept paths of constant temperatures for $
t \ge 0.81$ and constant $d$ for $d\ge 1.92$  and we have not found any
indication of the first-order line, separating the low temperature ordered
phases, terminating in the second order transition line. This is in
disagreement with previous renormalization group calculations \cite
{srg,s1.5mk}  and reproduce the qualitative behavior predicted by mean field
like calculations \cite{s1.5mf}, and are in agreement with our results 
of finite-sizes calculations on strips (see Fig. 2).
%

%
%
%
\newpage
\Large 
\bc 
Figure and Table Captions 
\ec
\normalsize
\vspace*{1.5cm}
\begin{figure}[h]
\caption{ Phase diagram in the ($d-t$)-plane for the spin-1 
Blume-Capel model. The continuum line is the second-order 
transition line and the dotted line is the first-order transition. 
The tricritical point is located at the end point of the continuum line. }
\label{Fig1}
\end{figure}
\begin{figure}[h]
\caption{ Phase diagram in the ($d-t$)-plane for the spin-3/2 
Blume-Capel model. The continuum line is the second-order 
transition line and the dotted line is the first-order transition. 
The circles were obtained by finite-size estimates and 
the squares by Monte Carlo.
 The full square represents the corresponding isolated multiphase critical
point. There are two different ferromagnetic phases with $m_1\rightarrow 3/2$, 
and $m_2\rightarrow 1/2$ as $T\rightarrow 0$. The open squares in the blown-up 
 region $1.90 \le d \le 2.10$ are obtained by Monte Carlo simulations.}
\label{Fig2}
\end{figure} 
\begin{figure}[h]
\caption{  Quadrupolar  susceptibility (open squares) and magnetization
           (open circles) as a function 
         of $t$ for  the Blume-Capel model with $S=3/2$
         and $d=1.998$. 
          $\chi _q $ has a strong peak only close 
         to $t=0.52$ and the magnetization a discontinuity
         of about $0.94$, indicating a first-order transition.
         The data were taken for increasing $t$.
         The lines are guide to the eyes.
         The data have been normalized: the maximum 
         magnetization is $m=1.499$ and the
         maximum susceptibility is $\chi_q=2.02$.}
\label{Fig3}
\end{figure}
\newpage
\begin{figure}[h]
\caption{  Magnetic  susceptibility (open squares) and magnetization
           (open circles) as a function 
         of $t$ for  the Blume-Capel model with $S=3/2$
         and $d=1.998$. 
          $\chi _m $ has a strong peak only close 
         to $t=0.66$ and the magnetization  is continuous, 
         indicating a second-order transition.
         The data were taken for increasing $t$.
         The lines are guide to the eyes.
         The data have been normalized: the maximum 
         magnetization is $m=1.499$ and the
         maximum susceptibility is $\chi_m=53.6$.}
\label{Fig4}
\end{figure}
%
%
%
%
%
\vspace{1.5cm}

\noindent Table 1 - Finite-size data $c^{N,N+1}$ given by (\ref{anoest})
 and extrapolations for  the conformal anomaly for some values of $d$. 
The  first four columns are the values for the $S=1$ model while the 
 last four are those of the $S=3/2$ model.

\vspace{1cm}

\noindent Table 2 - 
Finite-size data $x^N(1,-1)$ given by (\ref{dimest})
 and extrapolations for the scaling dimension of the parameter order
 for some values of $d$. 
The  first four columns are the values for the $S=1$ model while the 
 last four are those of the $S=3/2$ model.
\newpage
\Large
\bc
Table 1
\ec
\normalsize
%
\scriptsize
\btb[h]
\bc
\bt{|c||l|l|l|l||l|l|l|l||}                     \hline
             &  \mc{4}{c||}{S=1} & \mc{4}{c||}{S=3/2}          \\ \cline{2-9}
 $ _{\makebox[.5cm]{ $N$}} $  &$d$=-0.5 &$d$=0 &$d$=1.9 &$d$=1.9655 &$d$=-0.5 &$d$=0 &$d$=0.5 &$d$=2 \\ \cline{2-9}
 &$t_c$=1.794(7)&$t_c$=1.681(5)&$t_c$=0.764(7)&$t_c$=0.609(4)&$t_c$=3.549(8)&$t_c$=3.287(2)&$t_c$=2.972(3)&$t_c$=0.64(7)\\ \hline\hline\hline
4  &0.460931 &0.461723 &0.570695 &0.641586 &0.461970 &0.463160 &0.465282 &0.626807   \\ \hline
5  &0.478008 &0.478429 &0.578328 &0.666641 &0.478574 &0.479283 &0.480650 &0.611765   \\ \hline
6  &0.486062 &0.486307 &0.575809 &0.678638 &0.486400 &0.486858 &0.487798 &0.589182   \\ \hline  
7  &0.490323 &0.490480 &0.569959 &0.685059 &0.490544 &0.490864 &0.491549 &0.568249   \\ \hline
8  &0.492838 &0.492948 &0.563332 &0.688878 &0.492994 &0.493232 &0.493754 &0.551328   \\ \hline 
9  &0.494458 &0.494539 &0.556880 &0.691349 &0.494573 &0.494758 &0.495170 &0.538405   \\ \hline
10 &0.495569 &0.495633 &0.550943 &0.693048 &0.495659 &0.495807 &0.496142 &0.528798   \\ \hline
11 &0.496370 &0.496421 &0.545616 &0.694269 &0.496440 &0.496562 &0.496841 &0.521749   \\ \hline 
12 &0.496968 &0.497009 &0.540896 &0.695175 & & & &                              \\ \cline{1-5} 
13 &0.497427 &0.497462 &0.536738 &0.695864 & & & &                             \\  \cline{1-5}
14 &0.497788 &0.497817 &0.533085 &0.696401 & & & &                              \\ \hline \hline
$\infty$ &0.499(9)&0.499(9)&0.50(3)&0.698(9)&0.499(8)&0.499(8)&0.499(8)&0.50(5)   \\  \hline
\et
\ec
\etb
%
%
%
\Large
\bc
Table 2
\ec
\normalsize
\scriptsize
%
\btb[h]
\bc
\bt{|c||c|c|c|c||c|c|c|c||}                     \hline
             &  \mc{4}{c||}{S=1} & \mc{4}{c||}{S=3/2}          \\ \cline{2-9}
 $ _{\makebox[.5cm]{ $N$}} \setminus^{ \makebox[.3cm]{$d$} }$  &-0.5 & 0 &1.9 & 1.9655 &-0.5 &0 &0.5 &2 \\ \hline\hline
4  &0.118424 &0.118238 &0.100130 &0.071623 &0.118424 &0.117745 &0.117610 &0.089071   \\ \hline
5  &0.121069 &0.120923 &0.105237 &0.073095 &0.121185 &0.120449 &0.120484 &0.100904   \\ \hline
6  &0.122389 &0.122278 &0.108791 &0.073783 &0.122612 &0.121806 &0.121985 &0.108882   \\ \hline  
7  &0.123137 &0.123049 &0.111422 &0.074134 &0.123454 &0.122565 &0.122863 &0.113933   \\ \hline
8  &0.123602 &0.123532 &0.113439 &0.074315 &0.124003 &0.123022 &0.123424 &0.117077   \\ \hline 
9  &0.123913 &0.123855 &0.115022 &0.074400 &0.124391 &0.123312 &0.123807 &0.119047   \\ \hline
10 &0.124132 &0.124082 &0.116285 &0.074425 &0.124683 &0.123500 &0.124082 &0.120304   \\ \hline
11 &0.124292 &0.124248 &0.117305 &0.074409 &0.124913 &0.123625 &0.124287 &0.121121   \\ \hline 
12 &0.124413 &0.124373 &0.118136 &0.074362 &0.125102 &0.123706 &0.124446 &0.121662                              \\ \hline 
13 &0.124507 &0.124470 &0.118821 &0.074290 & & & &                             \\  \cline{1-5}
14 &0.124582 &0.124547 &0.119387 &0.074199 & & & &                              \\ \cline{1-5}
15 &0.124642 &0.124608 &0.119860  &0.074091 & & & &                              \\ \hline \hline
$\infty$ &0.1250(0)&0.1250(2)&0.12(2)&0.074(0)&0.13(0)&0.123(8)&0.125(5)&0.122(2)   \\  \hline
\et
\ec
\etb
%

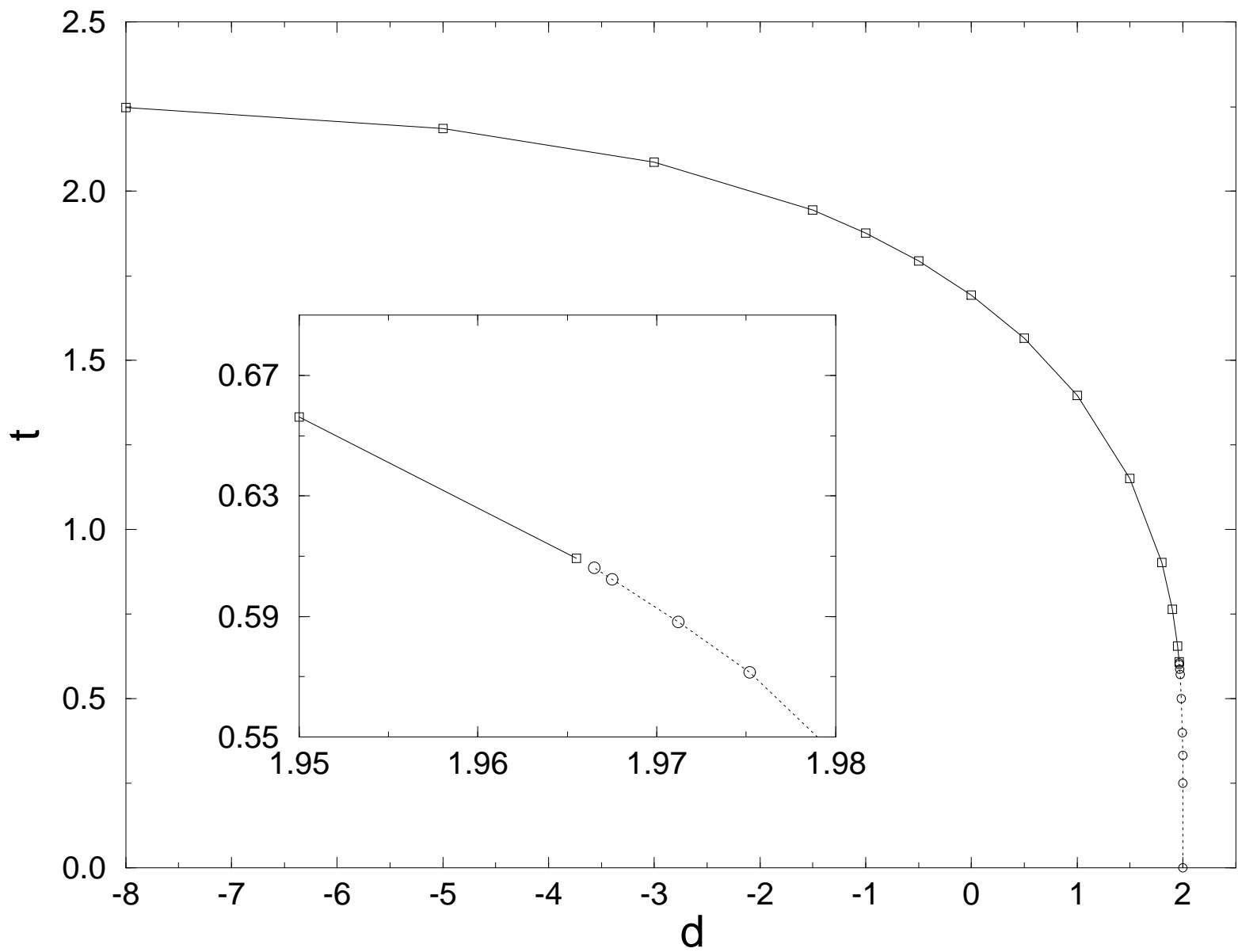
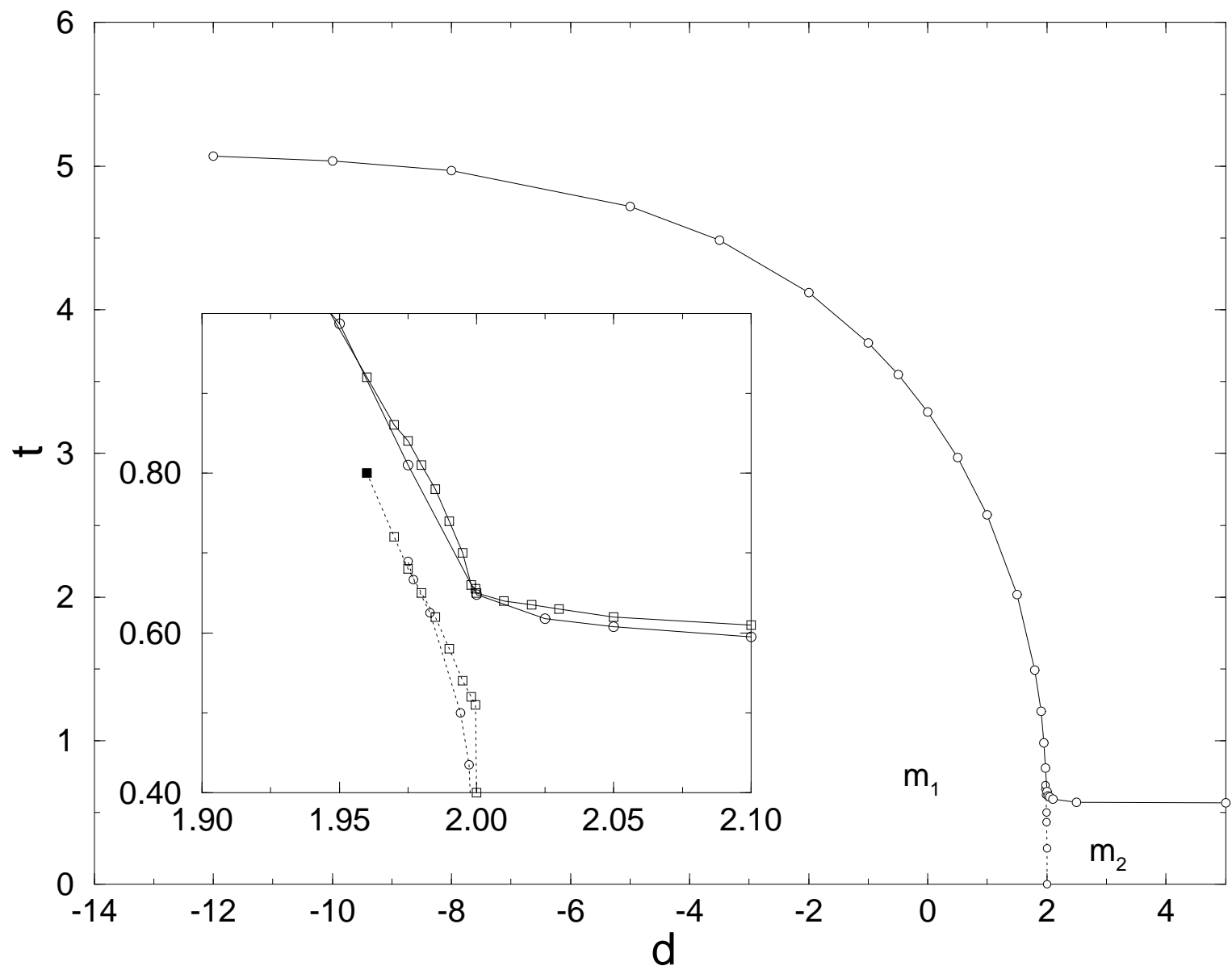
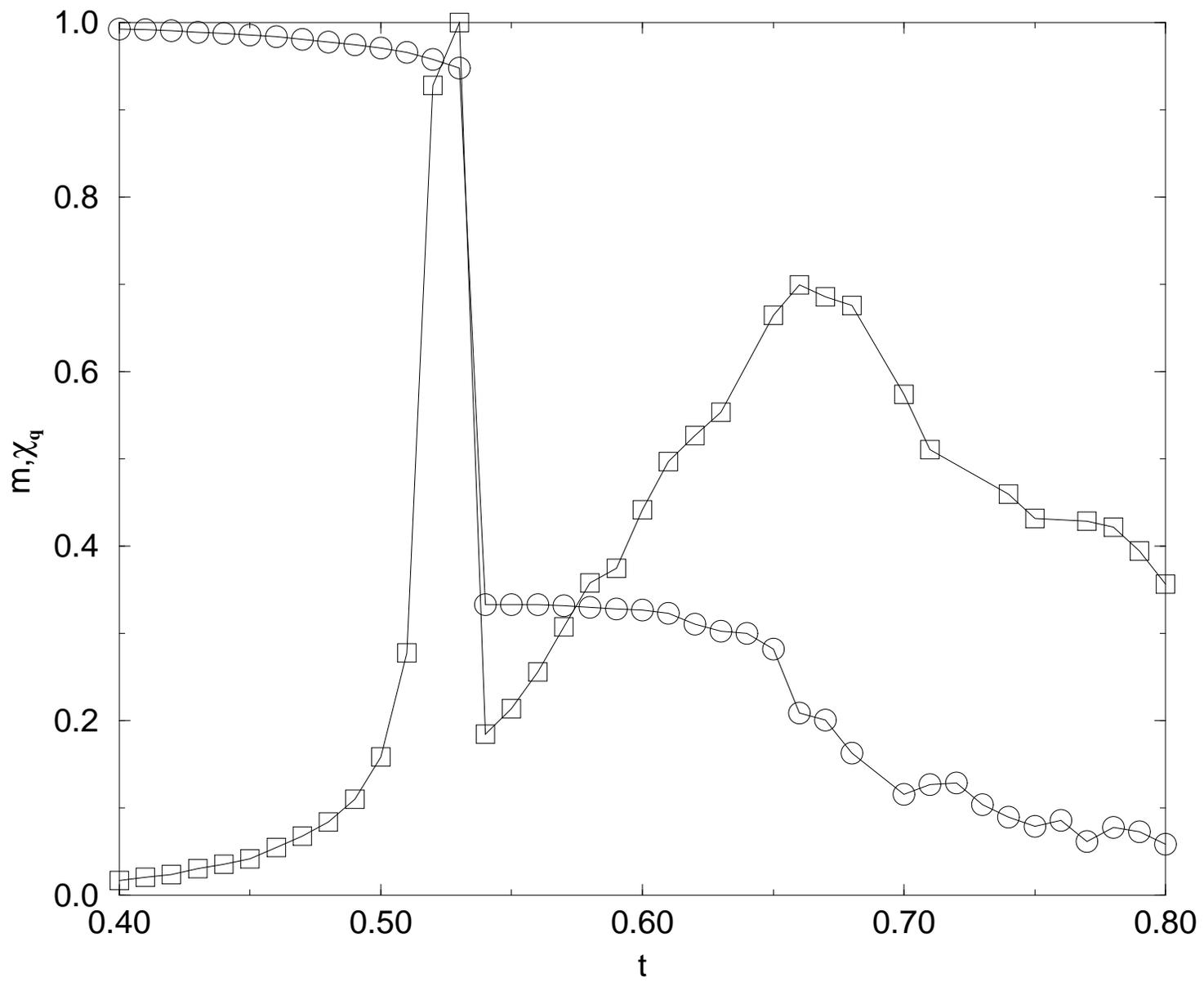
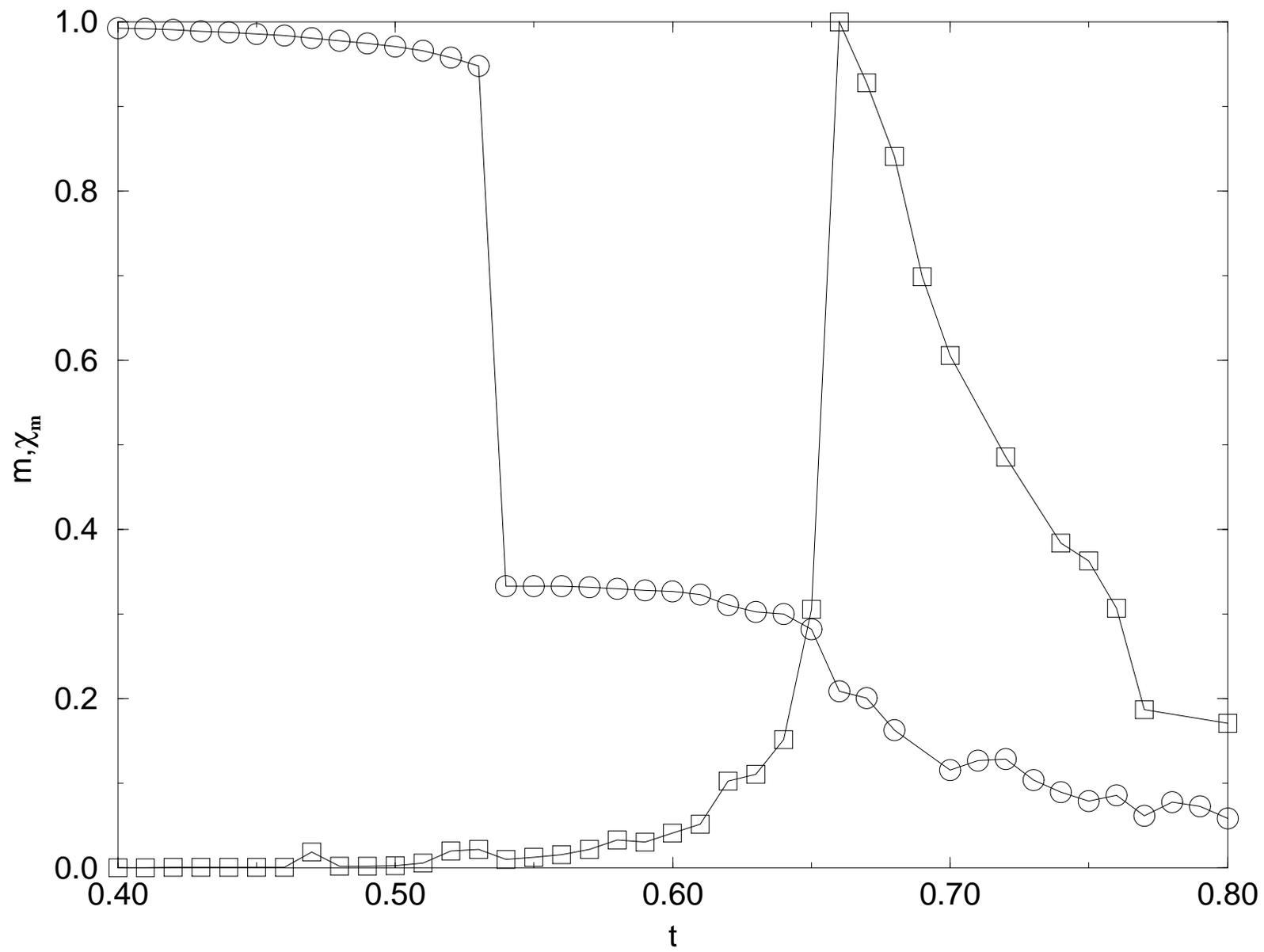
\end{document}